\documentclass{article}

\usepackage{graphicx,color}
\usepackage{amssymb}
\usepackage{amsmath}
\usepackage{epstopdf}
\usepackage[american]{babel}
\usepackage[T1]{fontenc}
\usepackage[utf8]{inputenc}
\usepackage{authblk}

\begin{document}

\title{Quantum cosmology with $k$-Essence theory}

\author[1]{C.R. Almeida\footnote{carlagbjj@hotmail.com}}

\author[1,2]{J.C. Fabris\footnote{fabris@pq.cnpq.br}}

\author[1]{F. Sbis\'a\footnote{fulviosbisa@gmail.com}}

\author[1,3,4]{Y. Tavakoli\footnote{ tavakoli@cosmo-ufes.org}}

\affil[1]{Departamento de F\'isica, UFES, Avenida Fernando Ferrari, 514, CEP 29075-910, Vit\'oria, ES, Brazil.}

\affil[2]{National Research Nuclear University ``MEPhI'',
	 Kashirskoe sh. 31, Moscow 115409, Russia}
	 
\affil[3]{Department of Physics, University of Tehran,  14395-547 Tehran, Iran}
\affil[4]{School of Physics, Institute for Research in Fundamental Sciences (IPM), 19395-5531 Tehran, Iran}

\vspace{.5cm}

\maketitle

\begin{abstract}
A class of $k$-Essence cosmological models, with a power law kinetic term, is quantised in the mini-superspace. It is shown that for a specific configuration, corresponding to a pressureless fluid, a Schr\"odinger-type equation is obtained with the scalar field playing the r\^ole of time. The resulting quantum scenario reveals a bounce, the classical behaviour being recovered asymptotically.
\end{abstract}

\mbox{Pacs:\,98.80.-k, 04.50.Kd}

\section{Introduction}

One of the main problems in the canonical quantisation of the Einstein-Hilbert Lagrangian \cite{c1,c2,c3,c4} is the absence of a clear time coordinate \cite{time1,time2}. There are many approaches to deal with this problem. One of them is to identify  an internal parameter that can play the r\^ole of time, a procedure called {\it deparametrisation} \cite{depa}. Another one, is to introduce matter fields such that they can be identified with the time coordinate. One example of the last procedure is to introduce a fluid with internal degrees of freedom using, e.g., Schutz's variables \cite{s1,s2}. In this case, the quantisation of the Einstein-Hilbert action coupled to a fluid, in the mini-superspace, leads to a Schr\"odinger-like equation, where the time coordinate is related to the conjugated momentum of the fluid variables, which appears linearly in the Hamiltonian \cite{ruba,demaret,lemos1,lemos2}.
The connection of the fluid variables with a time coordinate through Schutz's variable has been studied extensively in the literature. One interesting result is that the initial cosmological singularity is replaced by a bounce, and the classical solutions are recovered asymptotically \cite{lemos1,lemos2}. This scenario is consistent with the general belief that quantum effects must be important in the primordial universe, while our present universe is essentially classical.

Among the different proposals found in the literature to describe an accelerated phase of expansion of the universe, the $k$-Essence theories \cite{mukha1,mukha2,mukha3} have a very particular position. Conceived initially to describe the inflationary universe, the $k$-Essence theories have been used also to describe the present phase of accelerated expansion. This class of theories consider a non-canonical kinetic term instead of a self-interacting scalar field. In some cases, the $k$-essence behaviour can be recovered from an effective string action, as it happens with the DBI action \cite{dbi}. 
In a cosmological context, one of the characteristics of these theories is that, under some hypothesis, they can reproduce a fluid dynamics at background and perturbative levels \cite{scherrer,guio}. This is particularly true for the a kinetic power law expression, which can reproduce a linear relation between pressure and density $p = \omega \rho$, and the speed of sound for the adiabatic perturbations of the fluid.

In this letter we will investigate the possibility to obtain a time variable, in a way similar to the employment of Schutz's variables, using a power law non-canonical kinetic term. We will show that this is possible in a very special circumstance, which corresponds to a pressureless fluid. We will obtain a Schr\"odinger type equation, which will allow us to compute the expectation value for the scale factor, which reveals a bouncing universe in the same way as it occurs using the Schutz's variables. 

\section{A $k$-Essence quantum model}
\label{first}

The general $k$-Essence action can be written as \footnote{We use the signature $(+---)$ and the following convention for the Ricci tensor: $R_{\mu\nu} = \, \partial_\rho \Gamma^\rho_{\mu\nu} - \partial_\nu \Gamma^\rho_{\mu\rho}+ \Gamma^\rho_{\mu\nu}\Gamma^\sigma_{\sigma\rho} - \Gamma^\rho_{\mu\sigma}\Gamma^\sigma_{\nu\rho}$.} ,
\begin{eqnarray}
{\cal S} = \int dx^4\,\sqrt{-g}\Big\{R - f(X) + V(\phi)\Big\},
\label{action1}
\end{eqnarray}
where $g = \det g_{\mu\nu}$, and $f(X)$ is an arbitrary function of the kinetic term
$X \  =\  \phi_{;\rho}\phi^{;\rho}$
and $V(\phi)$ is a potential term. If $f(X) = X$, the usual minimally coupled system gravity-self interacting scalar field is recovered.

In what follows we will concentrate on the power law $k$-Essence model, for which $f(X) = \epsilon X^n$, where $n$ is a real number, and $\epsilon = \pm 1$. With the introduction of $\epsilon$ the possibility of a phantom configuration is taken into account. The usual gravity-scalar field system corresponds to $n = 1$, $\epsilon = 1$. Moreover, we will consider $V(\phi) = 0$. In this case, a cosmological fluid scenario with $p = \omega\rho$ and $\omega =$ constant is reproduced by the $k$-Essence model provided that
\begin{eqnarray}
\omega\  =\  \frac{1}{2n - 1}.
\end{eqnarray}
This particular $k$-Essence class of theories has been recently investigated in the context of static spherically symmetric configurations, revealing some very peculiar new structures \cite{denis}. 

Let us consider the flat, homogenous and isotropic Friedmann-Lema\^{\i}tre-Robertson-Walker (FLRW) metric defined by
\begin{eqnarray}
ds^2 = N(t)^2dt^2 - a(t)^2[dx^2 + dy^2 + dz^2],
\end{eqnarray}
where $N(t)$ is the lapse function.
With this metric, the action (\ref{action1}), after integrating by parts and discarding total derivatives, reduces to,
\begin{eqnarray}
{\cal S} = \int dt\,\biggr\{ \frac{6}{N}{\dot a}^2 a - \epsilon a^3N^{1 - 2n}{\dot\phi}^{2n}\biggl\}.
\end{eqnarray}
In order to have analicity, we will consider $\dot\phi$ positive, but it is possible to extend the results for the whole real line.
The corresponding conjugate momenta for the scale factor $a$ and the field $\phi$ are
\begin{eqnarray}
\pi_a = \frac{12}{N}a\dot a\quad ,\quad  \pi_\phi = - 2n\epsilon a^3 N^{1 - 2n}{\dot\phi}^{2n - 1}.
\end{eqnarray}
In expressing $\dot\phi$ in terms of $\pi_\phi$ we must invert the relation above. For $n = 2k$, $k$ is a natural number such that $k \neq 0$, the radicand must be positive ($\epsilon = - 1$); for $n = 2k + 1$, the radicand does not need to be positive, but analyticity is lost at the origin $\pi_\phi = 0$. In spite of this, we will proceed in a general way since the configurations that interest us imply different conditions on $n$.
The Hamiltonian reads $H \ =\  N{\cal H}$,
where
\begin{eqnarray}
{\cal H} \ =\  \frac{1}{24}\frac{\pi^2}{a} + (2n - 1)\,(-\epsilon a^3)^{-\frac{1}{2n - 1}}\biggr(\frac{\pi_\phi}{2n}\biggl)^\frac{2n}{2n - 1}.
\label{Hamiltonian}
\end{eqnarray}
If $n \rightarrow \infty$, the conjugated momentum associated to $\phi$ appears linearly in the Hamiltonian, so that, $\phi$ can play the r\^ole of time. 

\section{The case $n \rightarrow \infty$}
\label{quantum}

In the limit $n \rightarrow \infty$, the Hamiltonian takes the form,
\begin{eqnarray}
{\cal H}\ =\  \frac{1}{24}\frac{\pi_a^2}{a} + \pi_\phi.
\end{eqnarray}
A very important remark is that, even if the Hamiltonian is well defined in the limit $n \rightarrow \infty$, the Lagrangian is not well defined.
After the redefinition
$\frac{\phi}{24} \rightarrow \phi$, the corresponding Schr\"odinger equation, with $\hbar = 1$, reads
\begin{eqnarray}
- \partial^2_a\Psi - \frac{q}{a}\partial_a\Psi = a\,i\partial_\phi\Psi ,
\label{Schroedinger}
\end{eqnarray}
where we have introduced a factor ordering $q$.  This is essential the same equation found in reference \cite{lemos2} with the Schutz formalism. In what follows we will consider $q = 1$. In this case, it is possible to show that the effective Hamiltonian is self-adjoint \cite{moniz}. Other choices for $q$ could be used without changing in an essential way the final results.

The effective Hamiltonian represented by the terms in the left-hand-side of (\ref{Schroedinger}) is symmetric (or, hermitian) if a non-trivial measure is introduced in the computation of the scalar product:
\begin{eqnarray}
(\phi,\psi) \ =\ \int_0^\infty \phi^*\,\psi\,a^2\,da.
\end{eqnarray}
Let us consider a stationary state, such that
$\Psi(a,\phi) \  =\   \Phi(a)\,e^{-iE\phi}$.
Then, the Schr\"odinger equation (\ref{Schroedinger}) takes the form
\begin{eqnarray}
\partial^2_a\Phi + \frac{1}{a}\partial_a\Phi + a\,E\Phi = 0.
\end{eqnarray}
It is not difficult to show, using the non-trivial measure of the scalar product, that the energy is positive, $E > 0$, which is important for the stability of the system.
Changing to the variable $x = a^\frac{3}{2}$ and identifying, $\frac{4}{9}E \rightarrow E$, we end up with a Bessel's equation,
with the solution,
\begin{eqnarray}
\Psi(a,\phi) \  =\  A(E) J_0(E\,a^\frac{3}{2}) e^{-iE\phi},
\label{solution}
\end{eqnarray}
where $A(E)$ is a weight factor, and we have discarded the second solution of the Bessel equation, corresponding to the Neumann function, since it is divergent at the origin.

The solution (\ref{solution}) may lead to a non-singular cosmological scenario as in reference \cite{lemos2}. In fact, let us consider the wavepacket constructed with the following superposition \cite{grad}:
\begin{eqnarray}
\Psi_\phi(a) \  =\  \int_0^\infty y e^{-\alpha y^2} J_0(y\,a^\frac{3}{2}) dy \  =\  \frac{1}{2(\gamma + i\phi)}e^{- \frac{a^3}{4(\gamma + i\phi)}} ,
\end{eqnarray}
where $y = \sqrt{E}$ and $\alpha = \gamma + i\phi$, with $\gamma > 0$. 
Now,  we can calculate the expectation value for the scale factor, considering $\phi$ as the corresponding time variable. The expectation value is
\begin{eqnarray}
\langle a \rangle_\phi \ =\  \int_0^\infty \Psi^* a \Psi a^2\,da  \  =\  C(\gamma^2 + \phi^2)^\frac{1}{3},
\end{eqnarray}
where $C > 0$ is a constant. This implies a bouncing universe, with no singularity, since $\langle a \rangle_\phi \geq C \gamma^{2/3}$. Furthermore, asymptotically (that is when $\phi \rightarrow \infty$) we have $\langle a \rangle_\phi \propto \phi^\frac{2}{3}$.

We can easily verify that the corresponding classical cosmological scenario is recovered asymptotically. Using the FLRW, we find the differential equations (by fixing  the cosmic time, such that $N = 1$)
\begin{eqnarray}
\biggr(\frac{\dot a}{a}\biggl)^2 = \frac{(2n - 1)}{6}\epsilon{\dot\phi}^{2n}\quad , \quad
{\dot\phi}^{2n - 1} = Ka^{-3},
\end{eqnarray}
where $K$ is an integration constant.
Hence, we have the following equation for the scale factor:
\begin{eqnarray}
3\left(\frac{\dot a}{a}\right)^2\  =\   \bar K a^{-\frac{6n}{2n - 1}} \  =:\  \rho_\phi,
\end{eqnarray}
where $\bar K$ is a combination of the previous constants. A general solution can be easily obtained:
\begin{eqnarray}
a \propto t^\frac{2n - 1}{3n}\quad , \quad \phi \propto t^\frac{n - 1}{n}.
\end{eqnarray}
In the limit $n \rightarrow \infty$, the solutions read
\begin{eqnarray}
a \propto t^\frac{2}{3} \quad, \quad \phi \propto t.
\end{eqnarray}
The last relation confirms the previous statement that $\phi$ plays essentially the r\^ole of time in the limit $n \rightarrow \infty$. Moreover, in this limit, the scale factor behaves as in a dust dominated universe. We have classically the relation
$a \  \propto \  \phi^\frac{2}{3}$,
which agrees with the relation found asymptotically in the quantum model.

\section{Conclusions}
\label{conclusion}

In this letter we have studied a quantum model in the mini-superspace from a class of $k$-Essence cosmology based on a power law kinetic term, $X^n$, where $X$ is the usual expression for the kinetic term of a scalar field. We found that the momentum for the the scalar field appears linearly in the Hamiltonian in the limit $n \rightarrow \infty$. In this case, the scalar field may play the r\^ole of a time variable. The corresponding quantum scenario has been worked out, leading to a bounce universe, which recovers asymptotically the classical behaviour.
The case $n \rightarrow \infty$ leads, at classical level, to a cosmological model equivalent to that obtained by a pressureless fluid matter component, with $a \propto t^\frac{2}{3}$. A clear identification of the scalar field as time component is possible only in this special case. The canonical transformation allowing the identification of scalar field as a time component seems only well defined for that limit, otherwise we must face problems with fractional derivatives which may imply to loose the notion of locality. 
The fact that only the case corresponding to a pressureless fluid leads to a possible identification of the scalar field with a time variable evokes previous proposals that a pressureless fluid may allow to recover the notion of a time variable in cosmology \cite{dust,paw}.  It must be remarked, however, that strictly speaking, a pressureless fluid is an idealisation, since no real fluid has zero pressure exactly. In some sense, maybe such aspect of the problem is related to the curious properties of the original $k$-Essence model developed here in the limit $n \rightarrow \infty$, with a well defined Hamiltonian, but with no Lagrangian. The possible deep meaning of such limit process remains an open problem.

{\bf Acknowledgements:} We thank CNPq (Brazil) and FAPES (Brazil) for partial financial support. 
CRA and YT acknowledge support also from CAPES (Brazil). YT acknowledges partial financial support from INEF (Iran).

\end{document}